\documentclass{article}

\usepackage[utf8x]{inputenc}
\usepackage{polski}
\usepackage[polish]{babel}
\usepackage{listings}
\usepackage{color}

\usepackage[breaklinks,backref]{hyperref}
\hypersetup{colorlinks=true,linkcolor=black,citecolor=black,filecolor=black,urlcolor=black,breaklinks=true}

\title{C++11 - określanie typów}
\author{Piotr Beling \\ Uniwersytet Łódzki, Wydział Matematyki i Informatyki}
\date{\today}

\begin{document}
\maketitle

\begin{abstract}
Niniejszy artykuł jest jednym z~serii artykułów w~których zawarto przegląd nowych elementów języka C++ wprowadzonych przez standard ISO/IEC 14882:2011, znany pod~nazwą C++11.

W~artykule przedstawiono nowe możliwości związane z~określaniem typów zmiennych.
Opisano słowa kluczowe \lstinline|auto| i~\lstinline|decltype|, nową składnie deklarowania funkcji/metod oraz narzędzia zawarte w~pliku nagłówkowym \lstinline|<type_traits>|.
\end{abstract}

\tableofcontents

\cleardoublepage

\section{Wstęp}
Standard języka C++, znany jako C++11 (a także C++0x), został opublikowany we~wrześniu 2011 roku, w~dokumencie ISO/IEC 14882:2011\footnote{Dokument ten dostępny jest odpłatnie, jednakże ze strony \url{http://www.open-std.org} można pobrać jego darmową wersję roboczą \cite{CPP11STD}. Najnowsza (N3337) datowana jest na 16 stycznia 2012.}.

W~stosunku do~poprzedniego standardu (C++03), wprowadza on dużą liczbą nowych elementów, zarówno do samego języka, jak i do~jego standardowych bibliotek.

W~większość popularnych kompilatorów dość szybko zaimplementowano prawie wszystkie nowe elementy wprowadzone w~C++11.

Niniejszy artykuł, jeden z serii, stanowi krótki i~bogato zilustrowany przykładami przegląd nowych konstrukcji które wzbogaciły język.

Szczególną uwagę przywiązano w~nim do możliwość związanych ze~wskazywaniem typów zmiennych.

\section{AUTOmatyczna dedukcja typu}
Słowo kluczowe \lstinline|auto|, mało istotne w C++03, ma w~C++11 nowe znaczenie i~umożliwia m.in. dedukcję typu zmiennej na~podstawie typu wartości inicjującej:
\begin{lstlisting}
auto i = 5;
const auto c = 8;
\end{lstlisting}
oznacza to samo co:
\begin{lstlisting}
int i = 5;
const int c = 8;
\end{lstlisting}
gdyż \lstinline|5| oraz \lstinline|8| są typu \lstinline|int|.

Następujący kod (gdzie \lstinline|v| jest typu \lstinline|std::vector<int>|):
\begin{lstlisting}
for (std::vector<int>::iterator i = v.begin();
	 i != v.end();
	 ++i)
	//...
\end{lstlisting}
można w C++11 zapisać krócej:
\begin{lstlisting}
for (auto i = v.begin(); i != v.end(); ++i)
	//...
\end{lstlisting}

Jeśli po prawej stronie operatora przypisania pojawi się referencja, \lstinline|auto| będzie domyślnie oznaczało typ pozbawiony tej referencji. Referencję, jak i~inne modyfikatory, można jednak dopisać:
\begin{lstlisting}
int& f();
auto i = f();        // i jest typu int
auto& r = f();       // r jest typu int&
const auto c = f();  // c jest typu const int
\end{lstlisting}

\section{Obliczanie typu wyrażenia za pomocą \lstinline|decltype|}

Dzięki operatorowi \lstinline|decltype| możliwe jest użycie typu podanego wyrażenia, np.
\begin{lstlisting}
//int i;
decltype(i) j = i;	//j jest typu int
std::vector<decltype(i)> v(3, i);	//wektor z 3 kopiami i
decltype(v)::iterator iter;	//odpowiedni dla v iterator
\end{lstlisting}

Samo wyrażenie wewnątrz \lstinline|decltype| nie jest wartościowane i~w~związku z~tym nie powoduje żadnych efektów ubocznych, np.:
\begin{lstlisting}
int i = 1;
decltype(i++) j;	  // j jest typu int
assert(i == 1);
\end{lstlisting}

Zachowanie \lstinline|decltype| zostało tak określone, by pokrywało się z~intuicją większości programistów, jednocześnie dając jak największe możliwości. Dokładana definicja typu \lstinline|decltype(e)| jest jednak dość skomplikowana \cite{plwiki:decltype}:
\begin{itemize}
    \item jeśli wyrażenie \lstinline|e| jest odwołaniem do~zmiennej w~zakresie lokalnym lub w~przestrzeni nazw, statycznej zmiennej w~klasie lub parametru funkcji, to~wynikiem jest zadeklarowany typ zmiennej lub parametru, np.
    \begin{lstlisting}
		int i;
		struct A { double x; };
		const A* a = new A();
		decltype(i) x1; 		// typ to int
		decltype(a->x) x2; 	// typ to double
	\end{lstlisting}
    \item jeśli \lstinline|e| jest wywołaniem funkcji lub użyciem przeciążonego operatora, \lstinline|decltype(e)| oznacza zadeklarowany typ wyniku tej funkcji, kontynuując przykład:
    \begin{lstlisting}
		const int&& foo();
		decltype(foo()) x3; // typ to const int&&
	\end{lstlisting}
    \item w przeciwnym razie, jeśli \lstinline|e| to l-wartość, \lstinline|decltype(e)| to \lstinline|T&|, gdzie \lstinline|T| jest typem \lstinline|e|; jeśli \lstinline|e| jest r-wartością, to wynikiem jest \lstinline|T|, kontynuując przykład:
    \begin{lstlisting}
		decltype((a->x)) x5; // typ to const double&
	\end{lstlisting}
	Proszę zwrócić uwagę na~dodatkowe nawiasy które pojawiły się w~deklaracji \lstinline|x5| w~stosunku do \lstinline|x2|. Wymuszają one potraktowanie wyrażenia jako l-wartość.
\end{itemize}

%
%
%
%

\section{Nowa składnia deklarowania funkcji/metody}
W C++11 następujące deklaracje są tożsame:
\begin{lstlisting}
int f(int x, double y) { ... }	//C++03 i C++11
auto f(int x, double y) -> int { ... }	//tylko C++11
\end{lstlisting}

Nowa składnia deklarowania funkcji/metody została wprowadzona do C++11 by umożliwić wygodne wyliczenie typu zwracanego przez funkcję/metodę na~podstawie typów jej argumentów, np.:
\begin{lstlisting}
//Niech vec<T> będzie klasą reprezentującym wektor,
//mającą dwa pola typu T o nazwach x, y.
//Przykładowa implementacja operatora dodawania wektorów,
//z odpowiednim awansem typów:
template <typename A, typename B>
auto operator+(vec<A> a, vec<B> b) -> vec<decltype(a.x+b.x)> {
	return vec<decltype(a.x+b.x)>(a.x+b.x, a.y+b.y);
};
//...  przykład użycia:
vec<double> a(1.5, 2.7);
vec<int> b(1, 3);
auto c = a + b;	//c == vec<double>(2.5, 5.7);
\end{lstlisting}

Proszę zauważyć, że~następujący zapis nie jest prawidłowy gdyż wewnątrz \lstinline|decltype| argumenty \lstinline|a| i \lstinline|b| nie są jeszcze zdefiniowane:
\begin{lstlisting}
vec<decltype(a.x+b.x)> operator+(vec<A> a, vec<B> b) //...
\end{lstlisting}

Prawidłowy, choć mniej wygodny od pierwowzoru, jest za to zapis:
\begin{lstlisting}
#include <utility>	//dla std::declval
//...
template <typename A, typename B>
vec<decltype(std::declval<A>()+std::declval<B>())>
operator+(vec<A> a, vec<B> b) {
	return vec<decltype(a.x+b.x)>(a.x+b.x, a.y+b.y);
};
\end{lstlisting}
Zdefiniowany w~nagłówku \lstinline|utility| szablon \lstinline|std::declval| zwraca r-wartość typu podanego jako parametr szablonu, co umożliwia np. wskazanie pól tego typu. Typ ten może być niekompletny lub abstrakcyjny, zaś~jego obiekty niemożliwe do~utworzenia. Jednakże \lstinline|std::declval| można użyć jedynie w~operatorach które nie wartościują wyrażeń, takich jak \lstinline|decltype| czy \lstinline|sizeof|.

\section{Nagłówek \lstinline|<type_traits>|}

Nagłówek \lstinline|<type_traits>| zawiera zestaw narzędzi, głównie szablonów klas, pozwalających sprawdzać własności typów oraz tworzyć nowe typy poprzez modyfikacje istniejących (np. usunięcie referencji czy modyfikatora \lstinline|const|).

Większość szablonów klas zdefiniowanych w~\lstinline|<type_traits>| specjalizowana jest jednym typem.

Szereg szablonów klas, których nazwy zaczynają się od \lstinline|is_| (rzadziej od \lstinline|has_|) posiada jedno statyczne pole typu \lstinline|bool| o~nazwie \lstinline|value| które ustawiane jest na~\lstinline|true| tylko gdy typ będący parametrem ma~daną własność. Na przykład:
\begin{lstlisting}
std::is_abstract<T>::value	// T jest klasą abstrakcyjną?
\end{lstlisting}

Są one szczególnie użyteczne z \lstinline|std::enable_if| do wybrania implementacji funkcji lub metody w~zależności od~własności typu.
Na przykład niżej zdefiniowany szablon funkcji kopiuje tablicę używając efektywnego \lstinline|memcpy| gdy tablica jest typów które można bezpiecznie skopiować tą~metodą (mają trywialny operator kopiowania):
\begin{lstlisting}
template<typename T>
typename std::enable_if<
			std::is_trivially_copy_assignable<T>::value
>::type
mycopy(const T* source, T* dest, std::size_t count) {
    memcpy(dest, source, count*sizeof(T));
}
 
template<typename T>
typename std::enable_if<
			!std::is_trivially_copy_assignable<T>::value
>::type
mycopy(const T* source, T* dest, std::size_t count) {
    for(std::size_t i = 0; i < count; ++i)
        *dest++ = *source++;
}
\end{lstlisting}

Gdy pierwszym argumentem szablonu \lstinline|std::enable_if| jest \lstinline|true|, to zawiera on pole \lstinline|type| określające typ drugiego argumentu (domyślnie \lstinline|void|), w~przeciwnym razie jest on~pustą klasą.
W~ten sposób, dla dowolnego typu, tylko jedna z~wersji \lstinline|mycopy| będzie poprawnie zdefiniowana i~użyta (istnienie drugiej nie spowoduje błędu kompilacji\footnote{Ta zasada znana jest pod~nazwą SFINAE (Substitution failure is not an error).}).

Szablony klas których nazwy, typowo, zaczynają się od~\lstinline|add_|, \lstinline|remove_| oraz \lstinline|make_| udostępniają, w~polu \lstinline|type|, zmodyfikowaną wersję typu będącego argumentem. Na przykład:
\begin{lstlisting}
typename remove_const<T>::type	
// np. int gdy T jest const int
\end{lstlisting}

Szablon \lstinline|std::result_of| służy do~ustalenia wartości zwracanej przez  podany wskazany funktor:
\begin{lstlisting}
typename std::result_of<fun(int)>::type
// typ zwrócony przez fun dla argumentu typu int
\end{lstlisting}

\bibliographystyle{plplain}
\nocite{*}
\bibliography{cpp}

\begin{thebibliography}{1}

\bibitem{cppref}
C++ reference.
\newblock \url{http://en.cppreference.com}.
\newblock [dostęp: 2013-04-27].

\bibitem{enwiki:cpp11}
Wikipedia, the free encyclopedia: C++11.
\newblock \url{http://en.wikipedia.org/wiki/C++11}.
\newblock [dostęp: 2013-04-27].

\bibitem{plwiki:cpp11}
Wikipedia, wolna encyklopedia: C++11.
\newblock \url{http://pl.wikipedia.org/wiki/C++11}.
\newblock [dostęp: 2013-04-27].

\bibitem{plwiki:decltype}
Wikipedia, wolna encyklopedia: decltype.
\newblock \url{http://pl.wikipedia.org/wiki/Decltype}.
\newblock [dostęp: 2013-04-27].

\bibitem{CPP11STD}
ISO.
\newblock {\em ISO/IEC N3337=12-0027 Working Draft, Standard for Programming
  Language C++}.
\newblock International Organization for Standardization, 2012.

\end{thebibliography}

\end{document}